\documentclass[a4paper,11pt]{article}
\pdfoutput=1 % if your are submitting a pdflatex (i.e. if you have
\usepackage{jinstpub} % for details on the use of the package, please
\usepackage{soul}
\usepackage{url} 
\usepackage{hyperref}
\usepackage{graphicx}

\usepackage{ulem} % put a line crossing a letter

\usepackage{lineno}
\usepackage{multirow}
\usepackage{caption}
\usepackage{subcaption}
\title{Additive manufacturing of fine-granularity optically-isolated plastic scintillator elements}

\author[]{The 3DET collaboration,} % followed the example here: https://iopscience.iop.org/article/10.1088/1748-0221/6/07/P07005
\author[a,b,c]{S.~Berns}
\author[a,b,c]{E.~Boillat,}
\author[d]{A.~Boyarintsev,}
\author[e]{A.~De Roeck,}
\author[e]{S.~Dolan,}
\author[f]{A.~Gendotti,}
\author[d]{B.~Grynyov,}
\author[a,b,c]{S.~Hugon,}
\author[e,1,2]{U.~Kose,}
\author[d]{S.~Kovalchuk,}
\author[f]{B.~Li,}

\author[f]{A.~Rubbia,}
\author[d]{T.~Sibilieva,} 
\author[f]{D.~Sgalaberna,}

\author[f]{T.~Weber,}
\author[f]{J.~Wuthrich,}
\author[f]{X.~Y.~Zhao}

\affiliation[a]{Haute Ecole Sp\'ecialis\'ee de Suisse Occidentale (HES-SO), CH-2800 Del\'emont, Route de Moutier 14, Switzerland}
\affiliation[b]{Haute Ecole d'Ing\'enierie du canton de Vaud (HEIG-VD), CH-1401 Yverdon-les-Bains, Route de Cheseaux 1, Switzerland}
\affiliation[c]{COMATEC-AddiPole, CH-1450 Sainte-Croix, Technopole de Sainte-Croix, Rue du Progr\`es 31, Switzerland}
\affiliation[d]{Institute for Scintillation Materials NAS of Ukraine (ISMA), National Academy of Science of Ukraine (NAS), Nauki ave. 60, Kharkiv 61072, Ukraine}
\affiliation[e]{Experimental Physics department, European Organization for Nuclear Research (CERN), Esplanade des Particules 1, 1211 Geneva 23, Switzerland}
\affiliation[f]{Institute for Particle physics and Astrophysics, ETH Zurich, Otto-Stern-Weg 5, CH-8093 Zurich, Switzerland}

\emailAdd{umut.kose@cern.ch}
\note{Corresponding author}
\note{Now at the Institute for Particle physics and Astrophysics, ETH Zurich}

\abstract{
Plastic scintillator detectors 
are used in high energy physics as well as for diagnostic imaging in medicine, beam monitoring on hadron therapy, muon tomography, dosimetry and many security applications.
To combine particle tracking and calorimetry it is necessary to build detectors with three-dimensional granularity, i.e. small voxels of scintillator optically isolated from each other. 
Recently, the 3DET collaboration demonstrated the possibility to 3D print polystyrene-based scintillators with a light output performance close to that obtained with standard production methods. 
In this article, after providing a further characterization of the developed scintillators, we show the first matrix of plastic scintillator cubes optically separated by a white reflector material entirely 3D printed with fused deposition modeling.
This is a major milestone towards the 3D printing of the first real particle detector.
A discussion of the results as well as the next steps in the R\&D is also provided.
}


\begin{document}
\maketitle
\flushbottom

\section{Introduction}
\label{sec:intro}

Plastic scintillator detectors can be found in a wide range of scientific and industrial applications.
They are used for particle tracking and calorimetry in high energy physics, for diagnostic imaging in medicine, hadron therapy beam monitoring, dosimetry and many security applications. Plastic scintillators, due to their low cost and easy fabrication, provide an affordable approach to develop massive detectors with complex geometries and very good particle detection performance. 
Recent applications in neutrino oscillation experiments aim at combining a massive active target with three-dimensional (3D) particle tracking, precise calorimetry 
and sub-ns timing resolution; deploying a few million 
small fine-granularity optically-isolated plastic-scintillator voxels \cite{superfgd}. 
Other examples can be found in 
reactor neutrino experiments~\cite{solid}
or in the next-generation sampling calorimeters~\cite{calice}.

An obvious limitation is given by the challenge to produce and assemble many small cubes, e.g., 1~cm$^3$, with the required precision.
In particular,
when more complex geometries are needed, the traditional methods based, for example, on extrusion or injection moulding may become quite expensive
due to the required subtractive processes such as machining or drilling,
 and require a long production. 
An ideal solution would consist of 
fabricating all the individual plastic-scintillator voxels of the desired geometrical shape already assembled together and composing a single block of three-dimensionally segmented active material. 

Additive manufacturing (AM) techniques~\cite{AM_review} offer an intrinsically fast solution to develop complex objects with a high degree of geometrical freedom, and are becoming quite common in a wide range of applications in diverse industrial sectors, aerospace, construction,  and in the biomedical sector, using different material types including polymers, metal alloys, ceramics, glass, bio-materials, and composite materials.
Polymers can be 3D printed with solid-based, powder-based and liquid-based AM techniques. Fused Deposition Modelling (FDM)~\cite{AM_FDM} is the most conventional, cost-effective and, thus, widely used solid-based technique. 

The most common type of organic scintillator is composed of organic fluorescent compounds dissolved in a polymer matrix (polystyrene or polyvinyl-toluene).
Existing application of AM, using the state-of-the-art techniques, have successfully used a variety of polymer materials including polymer composites, thermoplastic polymers, polymer hybrids, etc. AM techniques have much inherent potential that makes such tooling a very attractive and promising technology for particle detection applications. 

Compared to other AM techniques, such as stereolitography (SLA), FDM allows the utilization of the very well-known optimal polystyrene-based scintillator and therefore does not necessitate the creation of a new chemical composition. This makes FDM a reliable approach for AM applications to scintillator-based particle detectors.

Recently, the 3D printed Detector (3DET) collaboration~\cite{3DETwebsite} was formed to investigate and develop AM as a new production technique for scintillators, 
and to perform a general-purpose R\&D toward the first 3D printed particle detector with performances comparable to the state of the art. 
The first proof of the concept showed the feasibility of 3D printing as a new production process of polystyrene-based plastic scintillator ~\cite{Berns:2020ehg}.

Among the near future goals of 
the 3DET collaboration is the 3D printing of a scintillator matrix composed of several optically-isolated plastic scintillator volumes.
This is considered as an example case study that will easily apply to future neutrino particle detectors as well as sampling calorimeters 
or neutron detectors. 
The collaboration started the R\&D study with FDM technology due to its versatility and cost effectiveness as well as rapid prototyping capabilities of specific shapes and patterns. The first phase of the R\&D program assessed the production of scintillating filament, the tuning of the 3D printing parameters and provided the first proof of the concept.

In this article we describe the current phase of the R\&D that aims to further characterize the 3D printed polystyrene-based scintillator, by measuring its attenuation length,
and producing a new white reflector filament optimized for FDM. 
The final goal is the 3D printing of a matrix of optically-isolated scintillator cubes.
In the following sections we will discuss the R\&D results obtained by the 3DET collaboration.

\section{Additive manufacturing of polystyrene-based scintillator}\label{rd3dprint}
3D printing plastic scintillator particle detectors by use of FDM technology requires scintillating and reflecting filaments with a stable size and material properties to feed through the rollers and nozzle. 
The optimal composition of the scintillating filament is obtained with polystyrene doped with 2\% by weight of p-terphenyl (pTP) and 0.05\% by weight of 2,2-p-phenilene-bis(5-pheniloxazole) (POPOP) and an addition of 5\% by weight of byphenil. Latter was used as a plasticiser to overcome filament breakage during its production and the printing process. This formula does not require the invention of a new chemical composition, since polystyrene is one of the most common polymers used in scintillator materials. 
The 3DET collaboration developed the entire process, including the production of the filament made of polystyrene scintillator and the tuning of the 3D printing parameters,
as detailed in~\cite{Berns:2020ehg}. 

\subsection{Transparency of the 3D-printed scintillator}

\begin{figure}[h]
\centering
{\includegraphics[width=7cm,height=5.5cm]{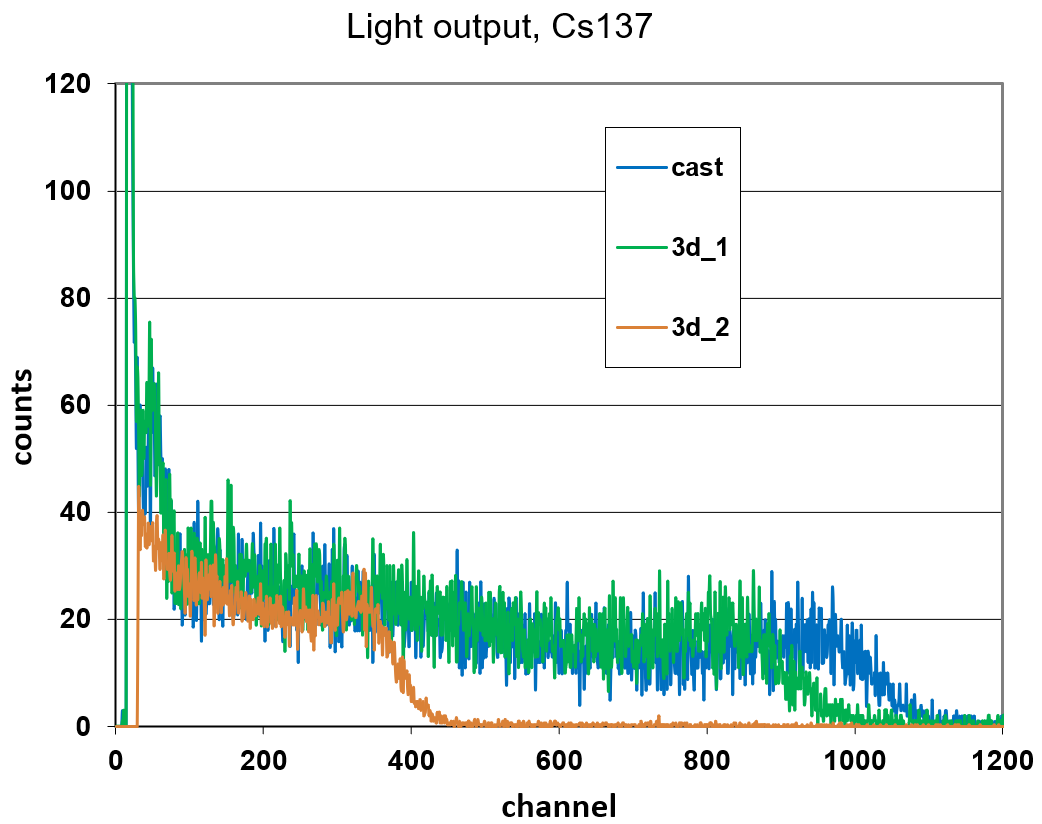}}
\caption{\label{fig:attenuationCs137} Light output from the 3D printed long bar sample with dimensions of about 10~mm$\times$10 ~mm$\times$50~mm exposed to a $^{137}$Cs $\gamma$-source is shown. The X axis shows the measured number of ADC channels. The comparison between cast scintillator and two 3D printed samples, obtained before (yellow) and after (green) the optimisation of the printing parameters is shown.}
\end{figure}

During the first phase of the R\&D program, samples of scintillator cubes with dimensions of 10$\times$10$\times$10~mm$^3$ were produced by using commercially available printers: “Roboze One+400”\cite{printer-roboze} and “CreatBot Dx2”~\cite{web:printer-creatbot}. The performance of the 3D printed scintillator was evaluated in terms of light yield and the number of photons produced per unit of energy deposited by charged particles~\cite{Berns:2020ehg}. 
An additional important characteristic
is the 
transparency to the scintillation light.
First measurements were performed by exposing each of the three scintillator bars made, respectively, with cast, extrusion and 3D printing methods to a  $^{137}$Cs $\gamma$ source. The measurement was performed by directly coupling scintillator samples to 3-inch PMTs (Hamamatsu R1307) using optical grease. The samples were irradiated by placing the source 10~mm away from the sample by help of cardboard spacer. The whole setup was then placed in a light tight black box covered with black blankets in order to prevent light coming from outside. A 150 seconds exposure time was used for data acquisition. The Compton edge was extracted from each measured spectra and used for comparison. 
As expected, the highest transparency was found in the cast scintillator sample. 
Figure~\ref{fig:attenuationCs137} illustrates the pulse height spectra for both the cast and the 3D printed samples, before and after optimizing the printing parameters.
A variation up to 50\% of the light output was found.
It was observed that the printing parameters, "bed and chamber temperature, fill factor", have to be carefully tuned in order to achieve the required transparency and light output. \\

After improving the printing parameters an acceptable attenuation length was obtained.
In order to quantify the impact of the 3D-printed scintillator transparency on the response to charged particles, measurements of the light attenuation were performed.
A 5-cm long scintillator bar with a cross section of $1\times 1~\text{cm}^2$ was printed with the optimised printing configuration.
It is important to clarify that, in order to precisely measure the bulk attenuation length, a sample with a length of at least tens of centimeters would be needed.
Hence, the aim is rather the measurement of a technical attenuation length, which depends on the geometrical configuration. This measurement was obtained after covering the scintillator bar with white teflon, which maximises the light internal reflection. 
This was found to be sufficient for broad assessment of the scintillator transparency and its implications on the response of few-cm granularity scintillator detectors.

\begin{figure}[h]
\centering
{\includegraphics[width=7cm, trim=0 -1.cm 0 0]{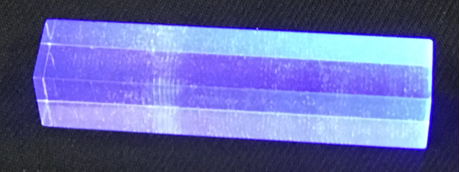}}
{\hspace{0.2cm}}
{\includegraphics[width=7cm]{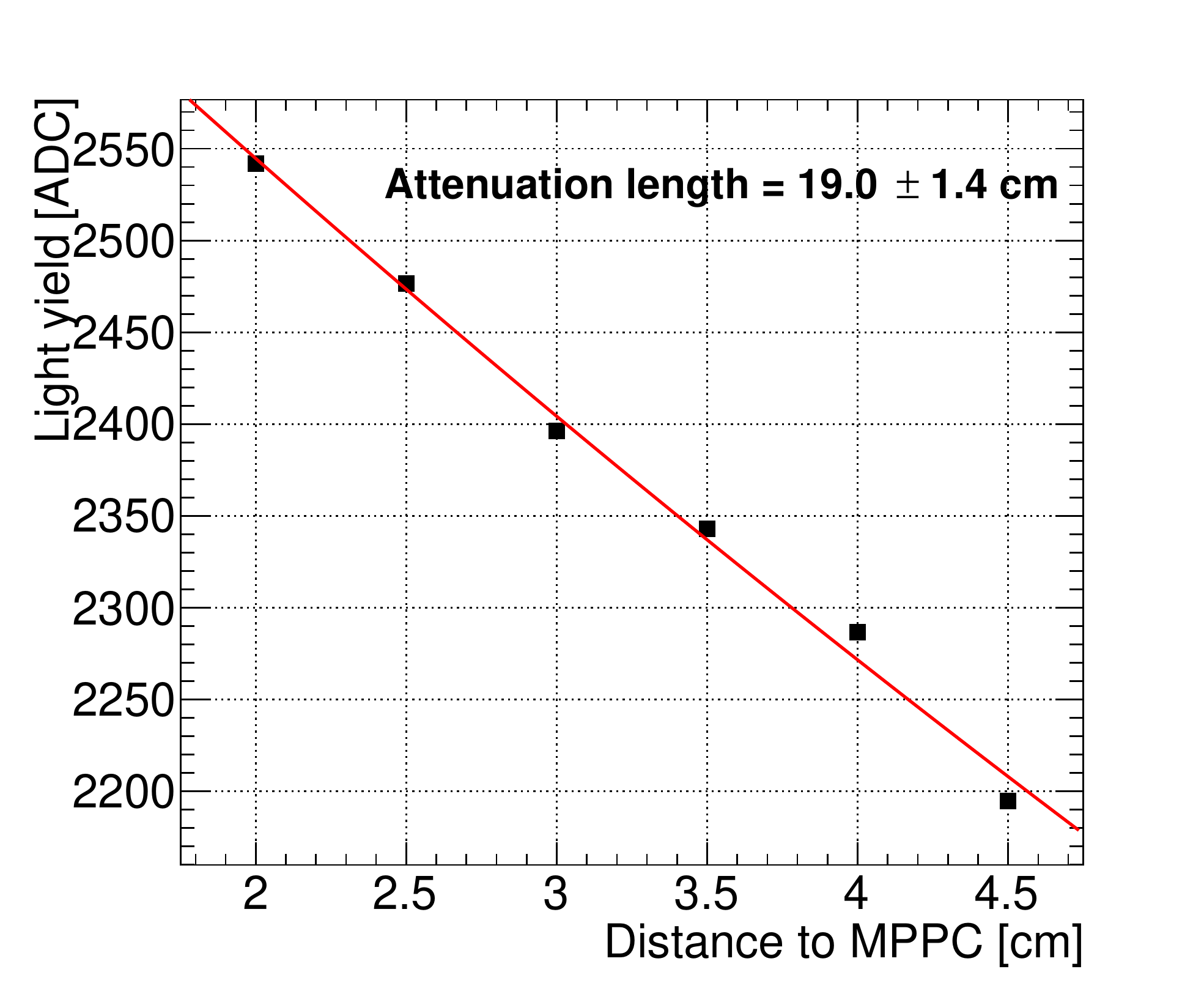}}
\caption{\label{fig:attenuationSr90} Left: the 3D printed scintillator bar with dimensions of $10~\text{mm} \times 10~\text{mm} \times~50~\text{mm}$ exposed to UV light. This sample is used to evaluate the attenuation length of the 3D printed scintillator. Right: the ADC threshold containing the highest 10,000 events as a function of the $^{90}$Sr source distance to SiPM is shown for measurements with one FEE threshold. The quoted uncertainty on the technical attenuation length comes from the measurements variation when using different FEE thresholds.}
\end{figure}

On the left panel of Figure~\ref{fig:attenuationSr90} the 3D printed scintillator bar is shown when it is exposed to ultraviolet (UV) light after the polishing of its outer surface. Whilst it can be observed that the bar is mostly transparent, very small air bubbles can be observed. Future improvements may be achieved by fine tuning the printing parameters in order to obtain a higher fill factor and, consequently, to remove the small air bubbles seen in the printed sample. 

The attenuation length of the 3D printed bar was measured by exposing it to a $^{90}$Sr source.
The scintillation light output was measured with a silicon photomultiplier (MPPC, Hamamatsu S13360-1350CS) directly coupled to one end of the bar. 
The light yield was measured for various distances between the $^{90}$Sr source and the SiPM. In order to ensure beta particles enter the bar at the same positions, a thin layer of paper indicating a central axis along the bar with step of half cm from far edge was placed between the source and the bar, without any gap, i.e. everything in contact.   
The setup was sensitive mainly to the energy spectrum region around the 1~MeV peak and above. However, it was found that inferring the position of the 1~MeV peak, which is relatively broad, can be affected by systematic uncertainties and, as a consequence, cannot provide a precise and accurate measurement of the technical attenuation length.
Instead a slightly different method, focused on the high-energy tail of the spectrum, was adopted.
First, in each measurement the scintillator bar was exposed to $^{90}$Sr for a fixed amount of time. The threshold of the front-end electronics (FEE) was set in a way to avoid any event loss. This ensured that the absolute integral of the spectrum tail is consistent among the different measurements and that enough events were collected to minimize the statistical uncertainty. 
Then, the ADC threshold that contains the 10,000 events in the highest ADC count part of the spectrum were obtained for each measurement and fitted, as shown in the right panel of Figure~\ref{fig:attenuationSr90}. 
Overall, a technical attenuation length of $19.0~\text{cm}$ was measured, which is acceptable for a finely segmented particle detector. It was additionally found that the measured attenuation length was not strongly affected by the choice of FEE threshold: the measurement varied by only $1.4~\text{cm}$ when considering a range of ADC thresholds between $\sim$2.5 p.e. and $\sim$6 p.e. without showing any particular trend.

\begin{figure}[h]
\centering
{
\includegraphics[width=12cm, trim=0 1000 300 1000,clip]{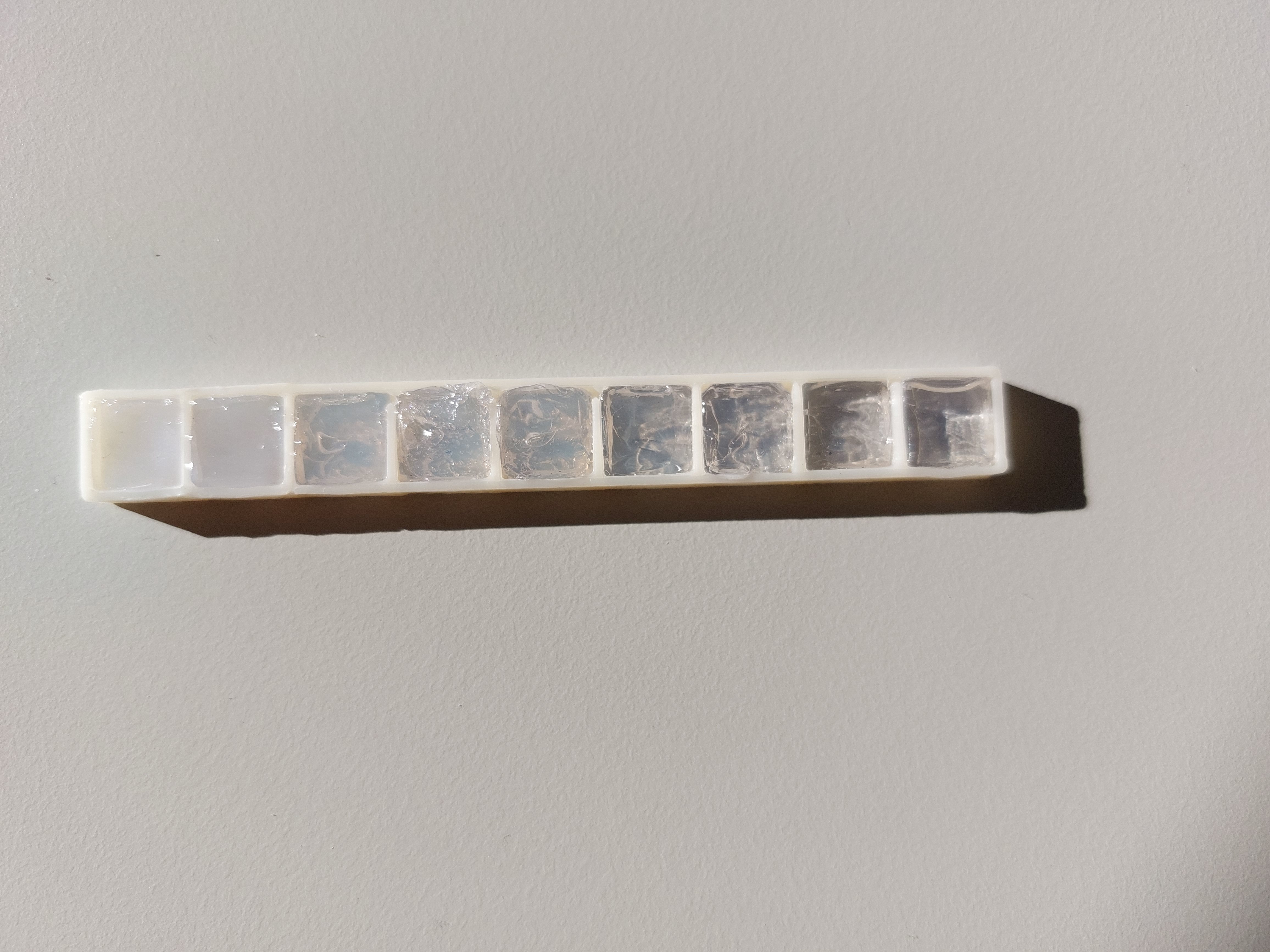} 
}
\caption{\label{fig:opaque-scintillator} 
Samples of 3D printed plastic scintillator obtained with different opacity to visible light.
}
\end{figure}

%%%%%%%%%%%%%%%%%%%%%%%%%%%%%%%%%%%%%%%%%%%%%%%%%%%%%%%%%
% Opaque scintillator
It is worth mentioning that optically-independent scintillator voxels, made out of very transparent scintillator covered by a white reflector layer is optimal for combined tracking, calorimetry and an accurate particle identification. Such a configuration avoids ambiguities introduced by the fact that the measured energy deposited in the scintillator voxel can depend on the position of the particle.  
The 3D printing of this configuration will be treated in detail in Sections~\ref{sec:reflector-filament}~and~\ref{sec:3dmatrix}. \\
However, an alternative solution suitable in particular for single-particle events 
can be achieved by confining the light in a reduced volume and detecting its radial spread with equally spaced wavelength-shifting fibers \cite{Tayloe:2006ct,Cabrera:2019kxi}.
In this context, we found that one can tweak the printing strategy in order to achieve different levels of opacity with a relatively good control, as shown in Figure~\ref{fig:opaque-scintillator}. This was obtained without introducing new additives to the scintillator filament, hence it is possible that the increased opacity is the result of a shortened scattering length rather than an altered absorption length.
Whilst a lot of efforts will be made in order to further improve the optical transparency of the 3D printed scintillator as well as the performance of the white reflective material, we also plan to characterize such features for potential alternative applications of the 3D printed scintillator.
%%%%%%%%%%%%%%%%%%%%%%%%%%%%%%%%%%%%%%%%%%%%%%%%%%%%%%%%%
%%%%%%%%%%%%%%%%%%%%%%%%%%%%%%%%%%%%%%%%%%%%%%%%%%%%%%%%%

\begin{figure}[h]
\centering
{\includegraphics[width=4.8cm,height=4cm]{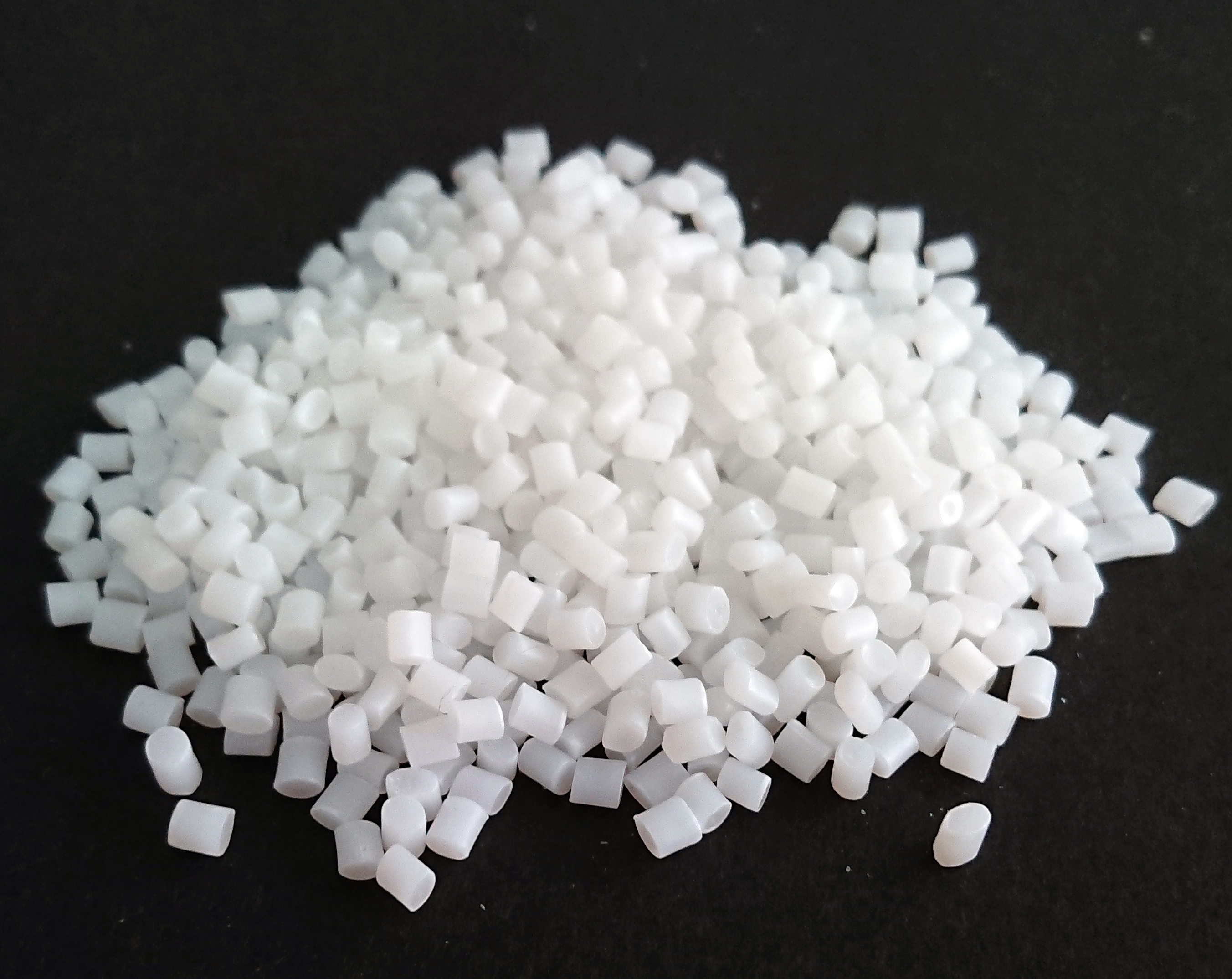}}
{\includegraphics[width=4.8cm,height=4cm]{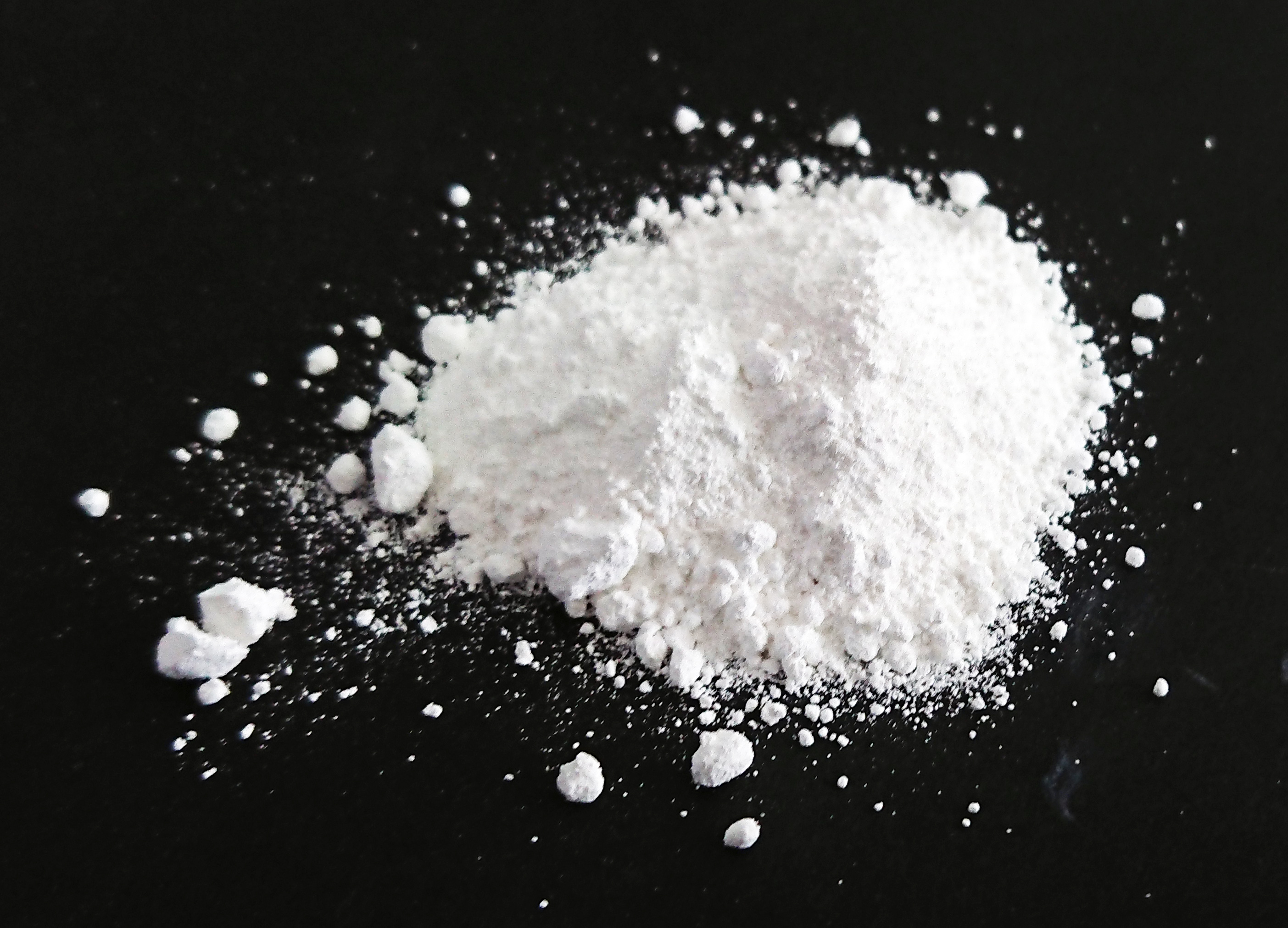}}
{\includegraphics[width=4.8cm,height=4cm]{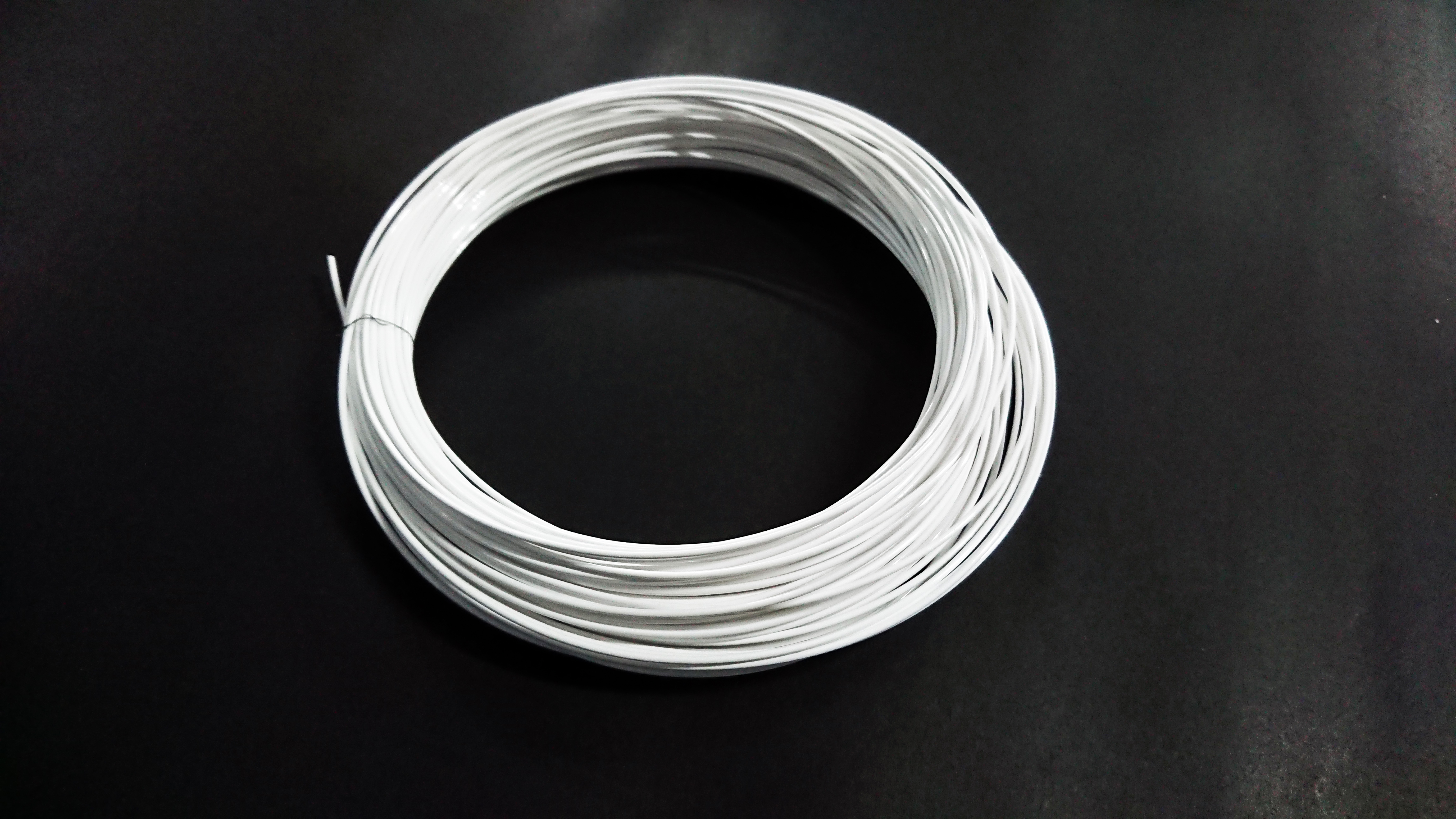}}
\caption{\label{fig:reflectivefilament} Reflective filament obtained by mixing polymer pellets (Left) with TiO$_2$ white reflective pigments (Middle) and then extruded to filament (Right).}
\end{figure}

\subsection{Optical reflector filament}
\label{sec:reflector-filament}

The function of the optical reflector filament is to segment the plastic scintillator into many optically-isolated three-dimensional voxels. 
The reflector filament can be obtained by dispersing commonly used reflective pigments, such as titanium dioxide (TiO$_2$), barium sulphate (BaSO$_4$), magnesium oxide (MgO), into polymer pellets like Acrylonitrile Butadiene Styrene (ABS), High Impact Polystyrene (HIPS), Polycarbonate (PC), Polymethylmethacrylate (PMMA) and Polystyrene (PS).
Then the polymer pellets mixed with the reflective pigments are extruded to obtain a filament, as seen in Figure~\ref{fig:reflectivefilament}. 
Different filaments were produced with 5, 10, 20 and 30\% reflective pigments in weight and used to 3D print samples of 20$\times$20~mm$^2$ with thickness between 0.15 and 1~mm. 
The reflectivity
spectrum was measured for each sample. 
The best results were obtained for those samples produced with 10 to 30\% TiO$_2$ reflective pigment mixed with PS or PMMA, as shown in Table~\ref{tab:reflectivity}. 
A reflectivity of 91\% at 420~nm was measured for a 1~mm thick reflector layer made of 20\% TiO$_2$ mixed with PS or PMMA . By optimizing printing parameters, such as the filling factor, a reflectivity of 92.5\% was reached. 

\begin{table}[h!]
\begin{center}
\begin{tabular}{ |c|c| } 
\hline
 Polymer mixed with $\text{TiO}_2$ & Reflectivity at $\lambda$=420 nm (\%)  \\
  \hline
  ABS  & 87.5 \\ 
 \hline
  HIPS  & 87.1 \\ 
 \hline
  PC  & 76.1 \\ 
 \hline
  PMMA  & 90.6 \\ 
 \hline
  PS  & 91.1 \\ 
 \hline
\end{tabular}
\caption{
\label{tab:reflectivity}
Reflectivity of FDM reflector filament samples with dimensions of $20\times20\times1~mm^3$, made of 20\% $\text{TiO}_2$ mixed with different polymer pellets. The same printing parameters were used in each case.
}
\end{center}
\end{table}

\begin{figure}[h]
\centering
{\includegraphics[width=7cm]{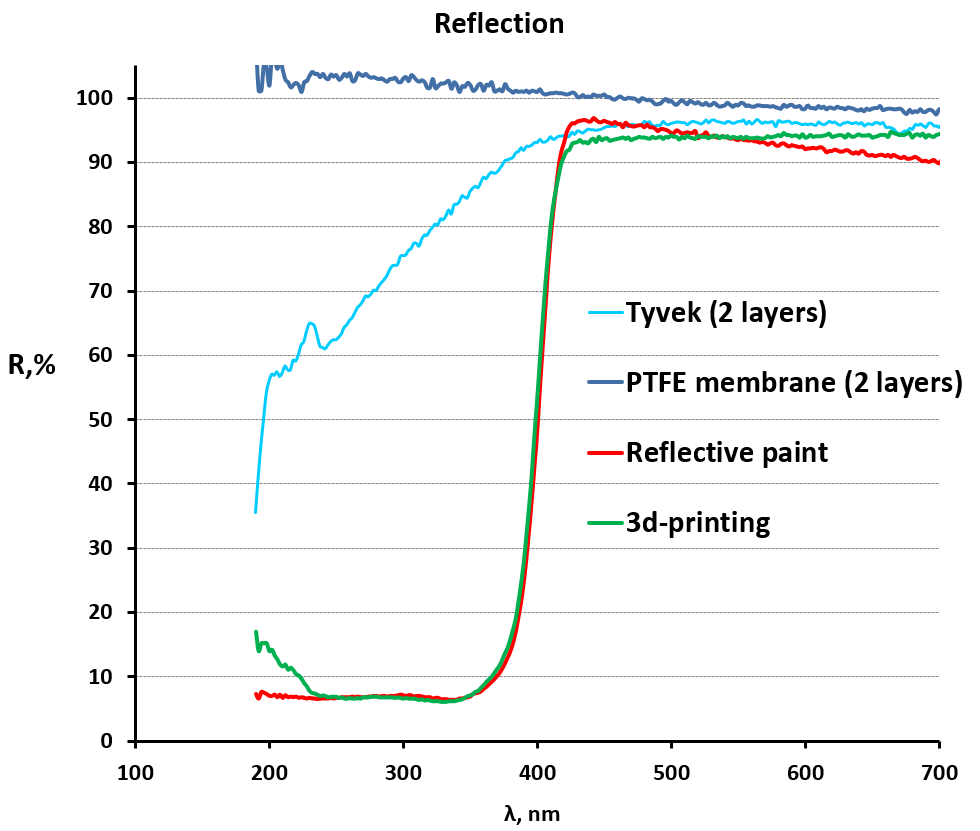}}
\caption{\label{fig:reflectivity} The reflectivity as a function of wavelength for different thicknesses of reflective materials: PTFE (0.34~mm), Tyvek (0.28~mm), TiO$_2$ reflective paint (0.15~mm) and the reflective filament optimized for 3D printing (0.40~mm).}
\end{figure}

Finally, a reflector filament with a diameter of 1.75~mm made of 20\% $\text{TiO}_2$ mixed with PMMA was produced and used for 3D printing optically isolated cubes. The reflectivity of the 3D-printed  TiO$_2$+PMMA filament was compared with different reflective materials, such as PTFE, Tyvek, and TiO$_2$ paint. We found similar reflectivity properties as Ti0$_2$ paint or tyvek at 420~nm for typical emission range of plastic scintillator as shown in Figure~\ref{fig:reflectivity}.

\section{Simultaneous printing of scintillator and optical reflector} 
\label{sec:3dmatrix}

\begin{figure}[h]
\centering
{\includegraphics[width=7cm,height=5.cm]{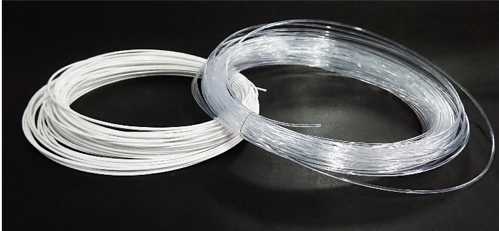}}
{\includegraphics[width=7cm,height=5.cm]{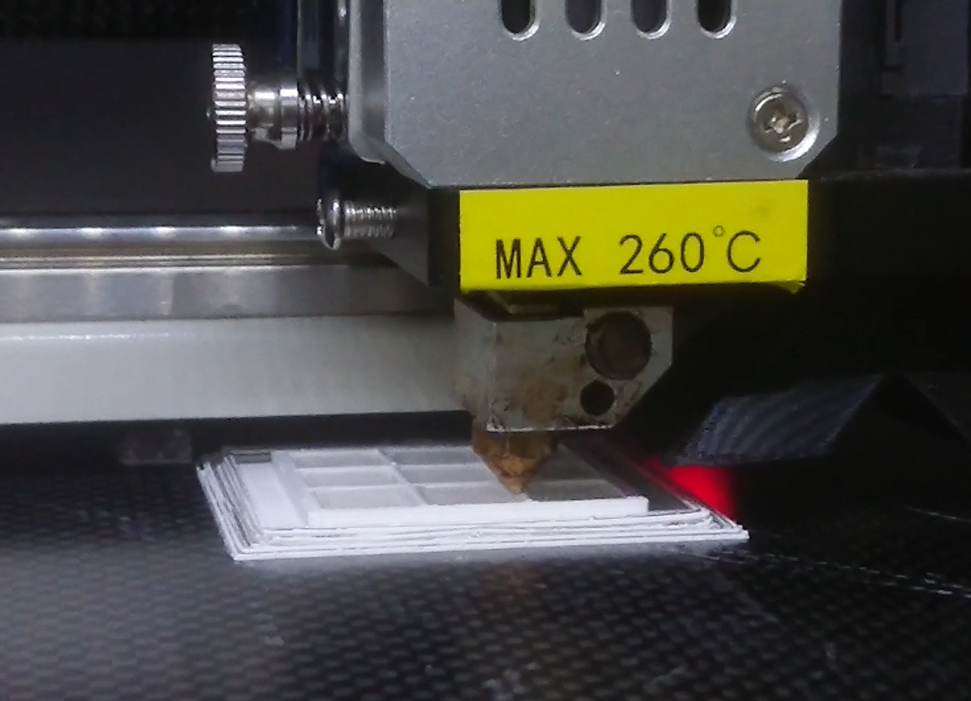}}
\caption{\label{fig:multiprint} Left: scintillating and reflecting filament used for 3D printing the cube matrix, Right: 3D~printing of 3$\times$3 cube matrix layer where the plastic scintillator cubes are optically separated by a 1~mm thick reflector wall.  
}
\end{figure}

In order to reach the final goal of 3D printing a single block of scintillator made of many optically-isolated cubes, the simultaneous printing of plastic scintillator and optical reflector was studied in detail. Following the method described in Section~\ref{rd3dprint}, scintillator and reflector filaments were produced as shown on the left panel in Figure~\ref{fig:multiprint}. Different samples were produced with the CreatBot F430 printer~\cite{web:printer-creatbot}, instrumented with two extruders as well as an air filter system. In the right panel of Figure~\ref{fig:multiprint} the multi-material 3D printing process is shown. 

\begin{figure}[h]
\centering
{\includegraphics[width=4.7cm,height=5.cm]{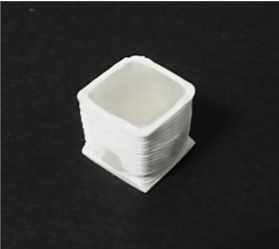}}
{\includegraphics[width=4.7cm,height=5.cm]{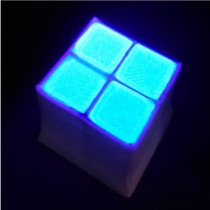}}
{\includegraphics[width=5.2cm,height=5.cm]{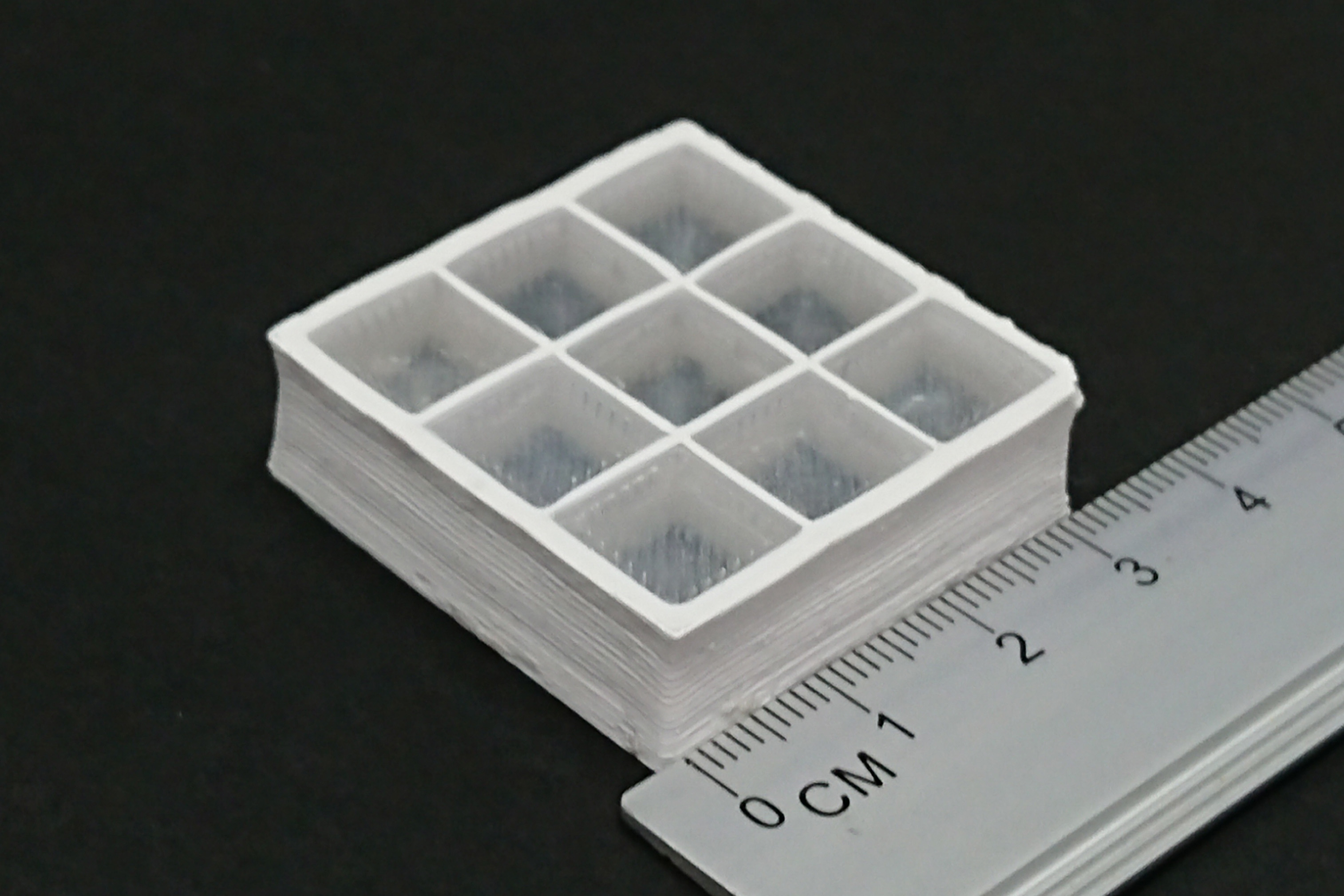}}
\caption{\label{fig:matrixlayers} Left: polystyrene-based scintillator cube with optical reflector. Middle: 2$\times$2 cube matrix layer exposed to UV light. Right: 3$\times$3 cube matrix layer. The plastic scintillator cubes are optically separated by 1~mm thick reflector.  
}
\end{figure}

The 3DET collaboration was successful at 3D printing a polystyrene scintillator cube with optical reflector and matrix of optically-isolated scintillator cubes.
Figure~\ref{fig:matrixlayers} shows samples of both cube and  matrices.
Each scintillator voxel corresponds to a 10~mm cube and $\sim$1~mm thick reflector. The geometrical precision was found to be acceptable for the inner part of the matrix within which the tolerance of the reflector thickness and cube shape were found to be $\sim$0.5 mm, while the outermost part of the matrix does not show a perfect rectangular shape. This is mainly due to the fact that the scintillator material needs to be melted to achieve an acceptable transparency and the outermost part is not mechanically constrained to preserve the structural accuracy of the matrix. However, this issue can be solved by post-processing the outermost surface if the required geometrical precision is not achieved already during the 3D printing stage. Moreover, some reflector remnants in the scintillator were observed since the extruder could not move up and down before changing the material. All this could be improved after further tuning of the 3D printing strategy. 

In the following sections the performance of the 3D~printed matrix is shown in terms of the scintillation output and optical crosstalk, the fraction of  scintillation light that leaks to an adjacent active voxel.

\subsection{Light output and light uniformity of the cube matrix layer}\label{sec:lightOuput}
\begin{figure}[h]
\centering
{\includegraphics[width=6cm,height=6.cm]{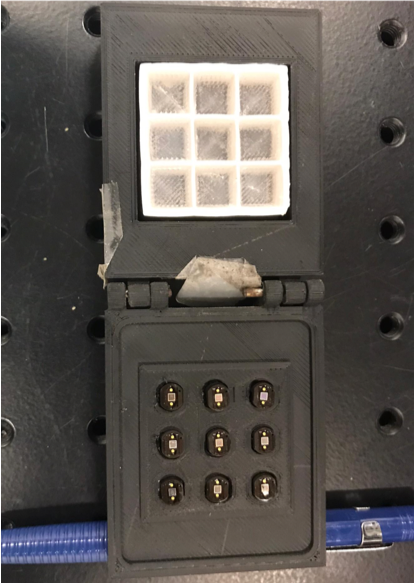}}
{\hspace{0.5cm}}
{\includegraphics[width=6cm,height=6.cm]{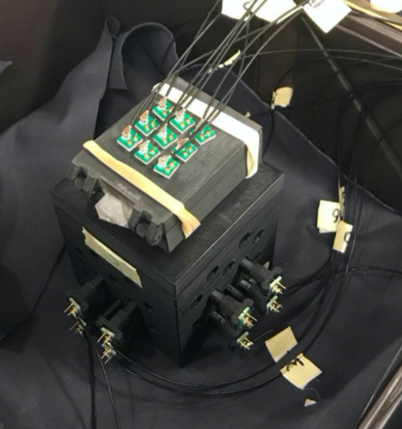}}
\caption{\label{fig:matrixtestsetup} Left: 3D~printed matrix layer setup, Right: setup used to selected vertical muons and evaluate the performance of 3D~printed matrix layer. The second box is hosting glued-cube layers read out by WLS fibers and MPPCs.  
}
\end{figure}

Measurements of the scintillation light output and light uniformity of the matrix layer were performed on a 3$\times$3 matrix sample. The scintillation light was collected by coupling a Hamamatsu S13360-1350CS Multi-Pixel Photon Counters (MPPC) directly to each scintillator cube of the matrix.
A 3D printed dark box, shown in Figure~\ref{fig:matrixtestsetup}, was used to provide the necessary light tightness. 
A piece of soft black EPDM foam was placed below each MPPC to push it against the cube, with the goal of improving the coupling with the scintillator. Teflon sheets were placed below and above the matrix layer, on the SiPM side 
some holes were made on the teflon sheet allowing coupling of the SiPM to scintillator cubes.  
Two layers of standard scintillator glued cubes
~\cite{Boyarintsev:2021uyw} 
read out with 1~mm diameter Kuraray Y11 single-cladding WLS fibers and MPPCs were also used to  select vertical muons. 
The MPPC charge signals were read out with a CAEN DT5702 front-end-board~\cite{web:caen}. The number of photoelectrons (p.e.) was extracted after measuring the MPPC gain, defined as the number of ADC counts per p.e.

\begin{figure}[h]
\centering
{\includegraphics[width=7cm,height=6.5cm]{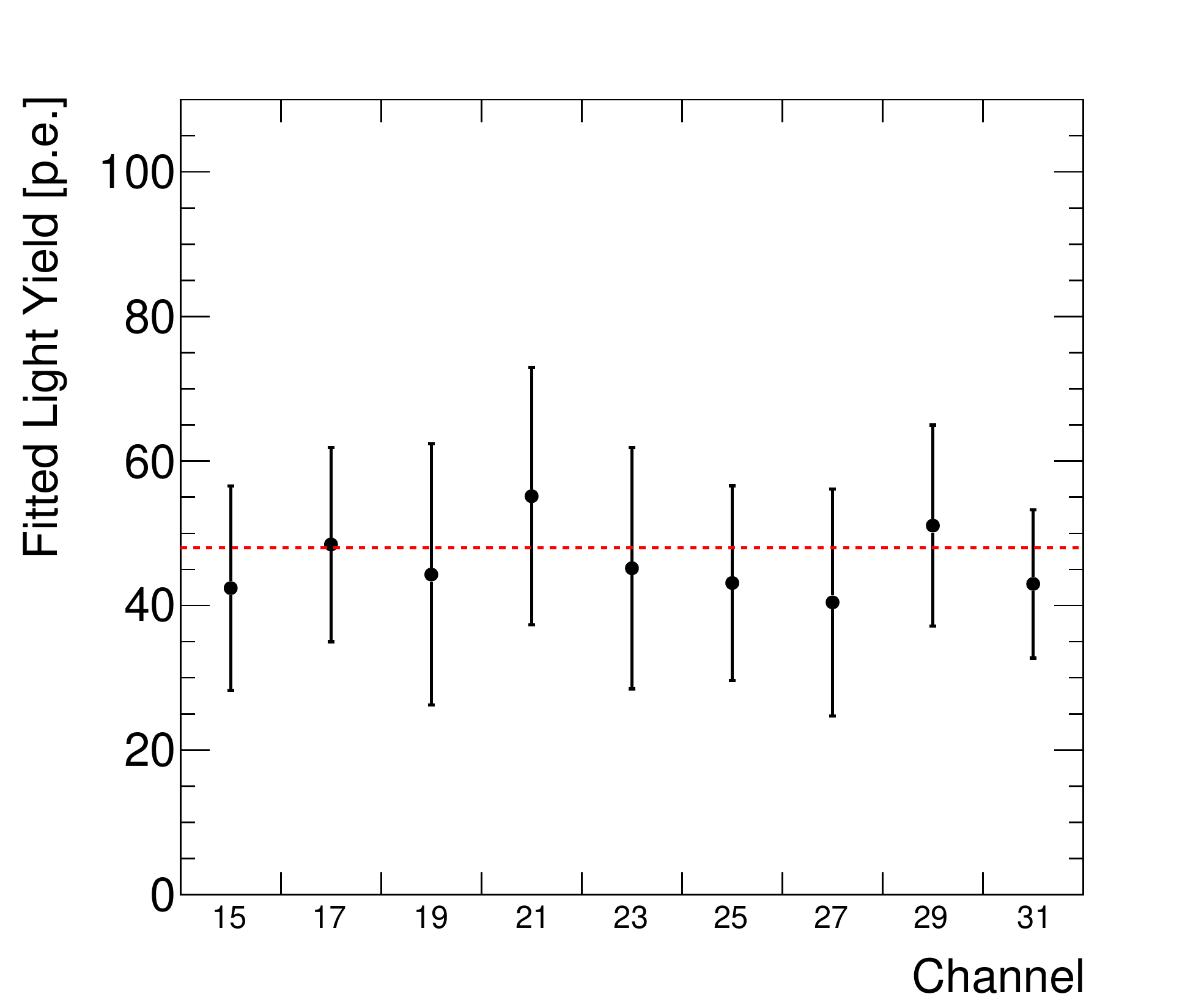}}
{\includegraphics[width=7cm,height=6.5cm]{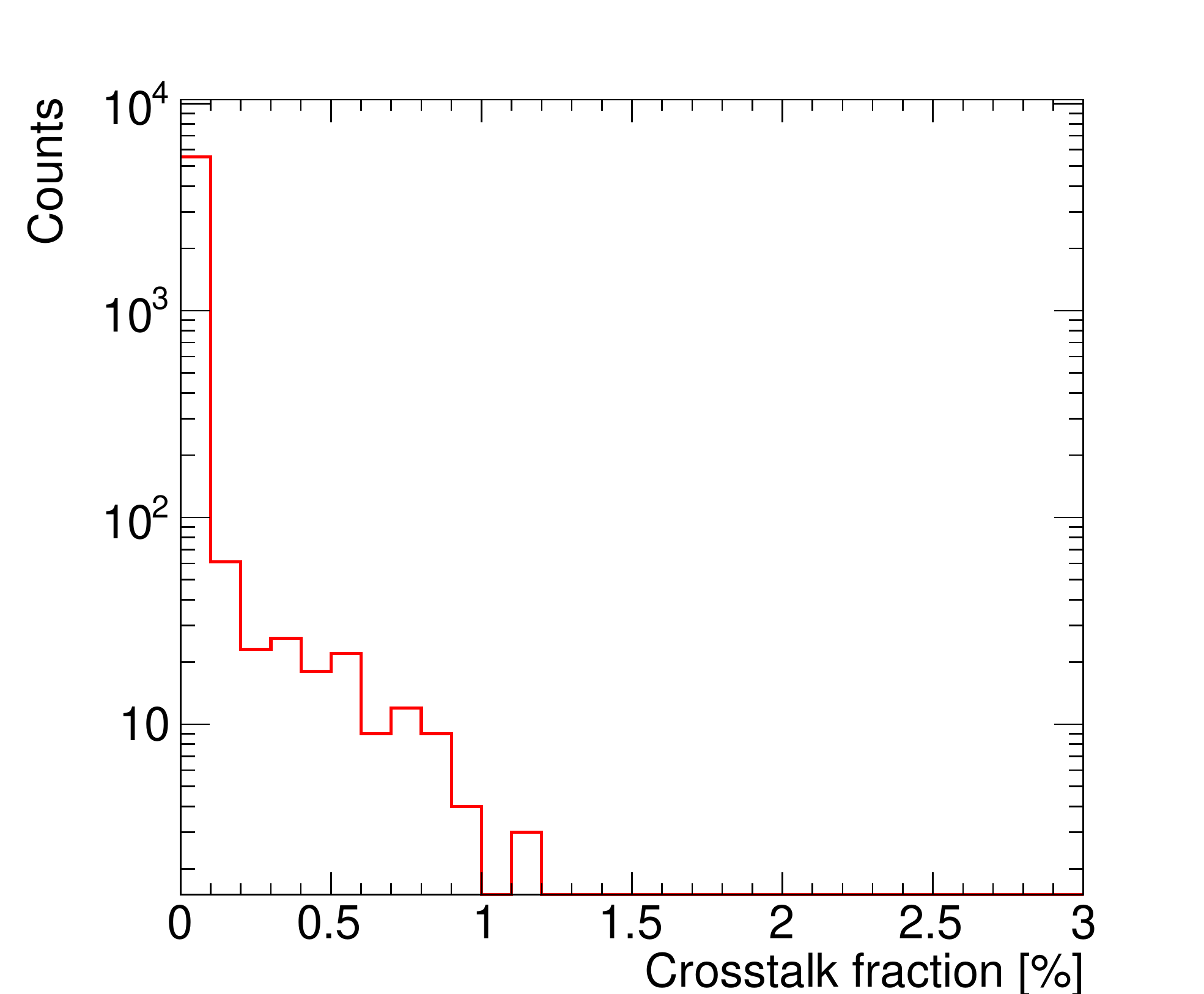}}
\caption{\label{fig:matrixtest}Left: average light output obtained by exposing the nine cubes of the 3D printed matrix to cosmics (vertical muons). Right: light cross talk between adjacent cubes from cosmic data taking. 
}
\end{figure}

The light yield of the cube matrix layer was measured from a sample of vertical cosmic muons, i.e. those crossing three layers on the same horizontal XY position. 
An ADC threshold was set to 400~ADC in order to minimise coincidences from background noise while rejecting almost no cosmic muons.
Before determining the light yield, a pedestal subtraction and gain correction was applied for each channel. The pedestal is established by taking the mean ADC count from events in which there is no hit above the threshold.  
First, the gain for each channel is determined using dedicated calibration runs, obtained by exposing the matrix layers to a $^{90}$Sr $\beta$ source. Once the pedestal and the gain are extracted for each channel, the light yield was measured by subtracting the pedestal from the observed number of ADC and then applying the gain correction.
The distribution of light yield obtained for each channel is fitted by a Gaussian function, allowing a determination of the most probable value and spread.
The average light yield from a cosmic muon is shown on the left panel of Figure~\ref{fig:matrixtest}. The light yield is found to be quite uniform among the different cubes. Approximately 45 p.e. are recorded by MPPC in a cube.

\subsection{Optical crosstalk measurement}
Optical crosstalk occurs when the light produced in a cube propagates to an adjacent cube. This could happen due to either the low diffusion capabilities of the reflector or a not-well-controlled thickness of the reflector layer.  
The crosstalk could interfere with the reconstruction process of the tracks, thus preventing from separating the signal contributions from different particles. Therefore, it is crucial for experimental purposes to make sure that the light crosstalk is kept low. 
The optical crosstalk was measured with vertical going cosmic muons following the same method described in Section~\ref{sec:lightOuput}. First the scintillation light signal produced by vertical muons was measured on each cube, called ``main cube'' and denoted by $M_{trk}$. Since the crosstalk is expected to take place only in neighboring cubes, named ``crosstalk cube'', their light signals, denoted by $M_{xtalk}$, were measured. Then, the crosstalk fraction was computed for each ``crosstalk cube'' as the ratio $M_{xtalk}/M_{trk}$.
As shown in the right panel of Figure~\ref{fig:matrixtest}, the optical crosstalk was found to be less than $\%$2 for the 3D~printed matrix layer, acceptable for applications that require a combined particle tracking and calorimetry.

\section{Conclusions}
\label{sec:conclusions}
The 3DET collaboration demonstrated the feasibility of 3D printing plastic scintillator detectors with Fused Deposition Modelling. The additive manufacturing process of a plastic scintillator prototype was improved by simultaneously printing both the scintillator and the optical reflector, after the tuning of the 3D printing parameters.
The 3DET collaboration was able to produce a 3D matrix of optically-isolated polystyrene scintillator voxels by additive manufacturing. 

More R\&D is planned to improve the geometrical tolerance and the transparency of the 3D printed scintillator. Further steps will aim to 
evaluate the 3D matrix reproducibility as well as the consistency of the printing performance.  
In order to obtain a full characterization of the scintillator, its decay time as well as the potential ageing effects will be studied. 

Overall, the results reported in this article are an important achievement of the ongoing R\&D program and a crucial milestone towards the realisation of the first 3D printed particle detector.  

Another potentially complementary feature is the tunable opacity of the 3D printed scintillator, that can be obtained without additional additive to the scintillator composition. More studies and tests will be done in order to obtain an accurate characterization.

%\bibliographystyle{unsrt}
%\bibliography{3Dmatrix}

%\iffalse

%\fi

\end{document}